# Crowd-Powered Data Mining

Chengliang Chai*, Ju Fan†, Guoliang Li*, Jiannan Wang‡, Yudian Zheng#

*Tsinghua University, †Renmin University, ‡SFU, #Twitter

## ABSTRACT

Many data mining tasks cannot be completely addressed by automated processes, such as sentiment analysis and image classification. Crowdsourcing is an effective way to harness the human cognitive ability to process these machine-hard tasks. Thanks to public crowdsourcing platforms, e.g., Amazon Mechanical Turk and CrowdFlower, we can easily involve hundreds of thousands of ordinary workers (i.e., the crowd) to address these machine-hard tasks. In this tutorial, we will survey and synthesize a wide spectrum of existing studies on crowd-powered data mining. We first give an overview of crowdsourcing, and then summarize the fundamental techniques, including quality control, cost control, and latency control, which must be considered in crowdsourced data mining. Next we review crowd-powered data mining operations, including classification, clustering, pattern mining, machine learning using the crowd (including deep learning, transfer learning and semi-supervised learning) and knowledge discovery. Finally, we provide the emerging challenges in crowdsourced data mining.

## 1 INTRODUCTION

Many data mining tasks cannot be effectively solved by existing machine-only algorithms, such as image classification [38], sentiment analysis [24], and opinion mining [7]. For example, given a set of pictures of famous places of interest in the world, we want to cluster them according to the country they belong to. Human can use their knowledge to categorize the pictures into countries like "China" or "America", but it is rather hard for machines. Fortunately, crowdsourcing has been emerged as an effective way to address such machine-hard tasks by utilizing hundreds of thousands of ordinary workers (i.e., the crowd). Thanks to the public crowdsourcing platforms, e.g., Amazon Mechanical Turk (AMT) and CrowdFlower, the access to the crowd becomes easier.

**Crowdsourcing Overview.** Over the past few years, crowdsourcing has become an active area from both research and industry (see a survey [32] and a book [31]). Typically, in a crowdsourcing platform (e.g., AMT [1]), there are two types of users, called *"workers"* and *"requesters"*. Requesters publish tasks on a crowdsourcing platform, while workers perform tasks and return the results. Suppose a requester has a classification problem to solve, which aims to classify an image to a category hierarchy. The requester needs to first perform *"task design"*, e.g., designing the user interface of a task (e.g., providing workers with an image and a category, and asking the workers to check whether the image belongs to the category), and set up some properties of the tasks (e.g., the price of a task, the number of workers to answer a task, the time duration to answer a task). Then the requester publishes the tasks to the platform. Workers who are willing to perform such tasks accept the tasks, answer them and submit the answers back to the platform. The platform collects the answers and reports them to the requester. If a worker has accomplished a task, the requester who publishes the task can approve or disapprove the worker's answers, and the approved workers will get paid from the requester.

**Challenges in Crowdsourcing.** The crowd has some different characteristics from machines. (1) *Not Free*. Workers need to be paid to answer a task, and it is important to control the *cost*. (2) *Error Prone*. Workers may return noisy results, and we need to tolerate the noises and improve the *quality*. Moreover, workers have various background knowledge, leading to different accuracies to answer different tasks. We need to capture workers' characteristics to achieve high *quality*. (3) *Dynamic*. Workers are not always online to answer tasks and we need to control the *latency*. Thus three core techniques must be considered in crowdsourcing: *"quality control"*, *"cost control"*, and *"latency control"*. Quality control aims to generate high-quality answers from workers' (possibly noisy) answers, by characterizing a worker's quality and aggregating workers' answers [2, 3, 8, 15, 21, 23, 25–28, 34, 40, 56, 63–65]. Cost control focuses on how to reduce human costs while still keeping good result quality [9, 13, 14, 21, 29, 37–39, 47, 49–51, 55, 59, 66, 67]. Latency control exploits how to reduce the latency by modeling workers' latency and estimating workers' arrival rates [18, 19, 22]. Note there are trade-offs among quality, cost, and latency, and existing studies focus on how to balance them, e.g., optimizing the quality given a fixed cost, reducing the latency given a fixed cost, minimizing the cost under latency and quality constraints, etc.

**Crowd Mining.** There are many studies that utilize the crowdsourcing to address the data mining tasks, including classification, clustering, patterning mining, outlier detection, and knowledge base construction and enrichment.

(1) *Crowd-Powered Pattern Mining*. Crowd pattern mining tries to learn and observe significant patterns based on workers' answers. The problem of discovering significant patterns in crowd's behavior is an important but challenging task. For example, a health researcher is interested in analyzing the performance of traditional medicine and she tries to discover the association rules such that "*Garlic can be used to treat flu*". In this case, she can neither count on a database which only contains symptoms and treatments for a particular disease, nor ask the healers for an exhaustive list of all the cases that have been treated. But social studies have shown that *although people cannot recall all their transactions, they can provide* *simple summaries (or called "personal rules") to the question*. For example, they may know that "*When I have flu, in most of the cases I will take Garlic because it indeed useful to me*". Given personal rules answered by different persons, they can be aggregated together to find an overall important rule (or the general trends). So the crowd pattern ming aims to collect the personal rules from crowd workers, aggregate them and find the overall important rules (i.e., general trends).

Crowd pattern mining typically generates a huge set of frequent patterns without providing enough information to interpret the meaning of the patterns. It would be helpful if we could also generate semantic annotations for the frequent patterns found, which would help us better understand the patterns. Existing works [4, 5, 46, 53, 57, 62] leverage the crowd ability to do semantic annotation, which mainly focus on improving the annotation accuracy and reducing the annotation cost.

(2) *Crowd-Powered Classification*. Some classification tasks are rather difficult for machines but easy for the crowd, e.g., image classification, and crowd-powered classification aims to leverage the crowd's intelligence to classify the data. Since the crowd may make mistake, existing works [6, 12, 35, 41–43, 54, 68] mainly focus on finding the correct classification from noisy crowd answers.

(3) *Crowd-Powered Clustering*. Many clustering tasks are easier for humans than machines. For example, given a set of sports events, human can easily categorize them into clusters like basketball, football according to their knowledge or experience. Some works focus on improving the clustering accuracy [20, 45, 52, 60]. Some works [11, 36, 48] not only care about the quality, but also optimize the cost or get high-quality results within a given budget.

(4) *Crowd-Powered Machine Learning*. Crowdsourcing can play an important role in machine learning, such as labeling data or debugging the model. There are several challenges of using the crowd in machine learning field. Firstly, when the number of data to be labeled bu human is very large, it is expensive for hiring either experts or the crowd. Therefore, we can utilize transfer learning or semi-supervised learning to do the task. Secondly, since the crowd workers are likely to make mistakes, we have to handle the errors. For example, deep learning can automatically tolerant the errors through the network. In this tutorial, we will discuss the usages of crowdsourcing in these advanced machine leaning algorithms in detail.

(5) *Knowledge Discovery*. We have witnessed the booming of large-scale and open-accessible knowledge bases (KBs), which contain thousands of millions of real-world entities, categories and relationships. However, despite the impressive size, no KB is complete. For example, KBs miss many entities, especially the long-tail entities. Thus, some existing works utilize crowdsourcing for knowledge base construction and enrichment, and existing studies can be classified into the following categories. (a) Crowd-powered knowledge acquisition: Kumar et al. [30] combine the crowdsourcing with information extraction techniques for knowledge acquisition in order to fill in missing relations among entities in KBs. (b) Crowd-powered entity collection: some works [10, 17, 44] utilize crowdsourcing to collect entities that are missing in a KB, e.g., collecting all active NBA players. (c) Crowd-powered knowledge integration: some works focus on integrating multiple KBs or linking entities in KB to external sources (e.g., web tables). For example, Zhuang et al. [69] leverage the crowd to align entities from multiple knowledge bases, which focuses on reducing the cost and achieving higher quality. Fan et al. [16] solicit the crowd to link categories in a KB to columns in web tables by using a hybrid human-machine approach. On the other hand, there are some works using knowledge base for better modeling the crowd. For instance, Zheng et al. [66] and Ma et al. [35] use the KB to model workers' quality considering the domain knowledge.

## 2 TUTORIAL AUDIENCE AND PREREQUISITE FOR THE TUTORIAL

This is a 3 hours' tutorial. The intended audience include all KDD attendees from both research and industry communities. We will not require any prior background knowledge in crowdsourcing.

## 3 TUTORIAL OUTLINE

We first give an overview of crowdsourcing, including motivation of crowdsourcing, basic concepts (e.g., workers, requesters, tasks), crowdsourcing platforms, crowdsourcing workflow, and crowdsourcing applications. Then we talk about fundamental techniques to address the three challenges: quality control, cost control, and latency control. Next, we discuss crowd mining operations. Finally we provide emerging challenges.

**Tutorial Structure:**

- Crowdsourcing Overview (20 minutes)
  – Crowdsourcing motivation
  – Crowdsourcing workflow
  – Crowdsourcing applications
  – Crowdsourcing platforms

- Quality control (40 minutes)
  Crowd workers may return relatively low-quality results or even noise. For example, a malicious worker may intentionally give wrong answers. Workers may have different levels of expertise, and an untrained worker may be incapable of accomplishing certain tasks. To achieve high quality, we need to tolerate crowd errors and infer high-quality results from noisy answers. The first step of quality control is to characterize a worker's quality (called worker modeling). For example, we can simply model a worker's quality as a probability, e.g., 0.8, i.e., the worker has a probability of 0.8 to correctly answer a task. To compute the probability, we can label some golden tasks with ground truth, and then the probability can be computed based on golden tasks. More sophisticated models will be introduced in our tutorial. Then based on the quality model of workers, there are several strategies to improve the quality in crowdsourcing. First, we can eliminate the low-quality workers (called worker elimination). For example, we can block the workers whose quality is below 0.6. Second, we can assign a task to multiple workers and infer the true answers by aggregating workers' results (called truth inference). For example, we can assign each task to five workers and then use majority voting to aggregate the answer. Third, we can assign tasks to appropriate workers that are good at such tasks (called task assignment). Thus, in order to build a robust, reliable and online crowdsourcing system, we need to (1) design smart truth inference algorithms that can tolerate crowd errors and infer high-quality results from noisy answers; (2) design online task assignment algorithms that can wisely use the budgets by dynamically assigning tasks to appropriate workers; (3) integrate truth inference and online task assignment in a self-learned system, by iteratively updating parameters (e.g., worker quality, task answers) based on workers' feedbacks and dynamically making reasonable online task assignment.

- Cost control (30 minutes)
  The crowd is not free, and if there are large numbers of tasks, crowdsourcing could be expensive. Thus cost control is indispensable in the system to prevent overspending. There are several effective cost-control techniques. The first is pruning, which first uses machine algorithms to remove some unnecessary tasks and then utilizes the crowd to answer the necessary tasks. For example, for image classification, we can prune the categories of an image where the image has a rather small possibility to belong to these categories. The second is task selection, which prioritizes the tasks and decides which tasks to crowdsource first. For example, suppose we assign a classification task to five workers iteratively. If three workers return that the image belongs to the given category, we do not need to ask the fourth and fifth workers. The third is answer deduction, which crowdsources a subset of tasks and deduces the results of other tasks based on the answers collected from the crowd. For example, for a dog image, if the crowd returns that it belongs to the category "Dog" and then we do not need to ask whether it belongs to "Animal". The fourth is sampling, which samples a subset of tasks to crowdsource. For example, suppose we want to know the percentage of "Dog" images in a dataset, we can use sampling to compute an approximate answer. The fifth is task design, which designs a better user interface for the tasks. For example, we do not want to enumerate every category in a hierarchy, and instead we can use a tree interface to ask the workers to easily identify the corresponding category.

- Break (30 minutes)

- Latency control (20 minutes)
  Crowd answers may incur excessive latency for several reasons: for example, workers may be distracted or unavailable, the tasks may not be appealing to enough workers, or the tasks might be difficult for most workers. If the requester has a time constraint, it is important to control latency. Note that the latency does not simply depend on the number of tasks and the average time spent on each task, because crowd workers perform tasks in parallel. Existing latency-control techniques can be classified into three categories. (1) *Single-task latency control* aims to reduce the latency of one task (e.g., the latency of labeling each individual image). (2) *Single-batch latency control* aims to reduce the latency of a batch of tasks (e.g., the latency of labeling 10 images at the same time). (3) *Multi-batch latency control* aims to reduce the latency of multiple batches of tasks (e.g., adopting an iterative workflow to label a group of images where each iteration labels a batch of 2 images).

- Crowd Mining (60 minutes)
  There are many data mining tasks can be achieved with higher quality by the crowd. In this tutorial, we will talk about pattern mining(10 min), classification (10 min), clustering(10 min), machine learning (10 min) and knowledge discovery(20 min). We will illustrate how to design crowdsourced tasks according to different data mining tasks, how to achieve high quality results and how to reduce cost and latency.

- Conclusion and future directions: (10 minutes)

## 4  TUTORS

**Chengliang Chai**
Affiliation: Tsinghua University
E-mail: chaicl15@mails.tsinghua.edu.cn
Address: Department of Computer Science, Tsinghua University, Beijing, China
Phone: 86-13001266011

**Ju Fan**
Affiliation: Renmin University of China
E-mail: fanj@ruc.edu.cn
Address: Room 500, Information Building, Renmin University of China, No.59 Zhongguancun Street, Beijing 100872, China
Phone: 86-10-62512304

**Guoliang Li (corresponding author)**
Affiliation: Tsinghua University
E-mail: liguoliang@tsinghua.edu.cn
Address: Department of Computer Science, Tsinghua University, Beijing, China
Phone: 86-10-62789150

**Jiannan Wang**
Affiliation: Simon Fraser University
E-mail: jnwang@sfu.ca
Address: School of Computing Science, Simon Fraser University, Burnaby, Canada
Phone: 1-7787824288

**Yudian Zheng**
Affiliation: Twitter Inc.
E-mail: yudianz@twitter.com
Address: 1355 Market St., Suite 900, San Francisco, CA 94103, US.
Phone: 1-4153702403

## 5  TUTOR'S BIO AND EXPERTISE

**Chengliang Chai** is a third year PhD candidate student at the Department of Computer Science, Tsinghua University, Beijing, China. He obtained his B.E. in 2015 from Harbin Institute of Technology, China. He visited University of Wisconsin-Madison in 2017. His research interests mainly include data cleaning and integration and crowdsourcing (especially cost control and latency control).

**Ju Fan** is currently working as an associate professor at the Department of Computer Science in Renmin University, Beijing, China. His main research interests include knowledge base and crowdsourcing (especially cost control and knowledge base construction and enrichment). He has published more than 30 papers in leading international conferences/journals, including SIGMOD, VLDB, ICDE, ICDM, etc. He got ACM China 2017 rising star award.

**Guoliang Li** is currently working as an associate professor at the Department of Computer Science, Tsinghua University, Beijing, China. His research interests mainly include data cleaning and integration, and crowdsourcing (especially cost control, quality control, latency control). He has published more than 100 papers in SIGMOD, KDD, VLDB, ICDE, SIGIR, VLDB Journal, TODS, TKDE, and his papers have been cited by more than 5000 times. He got VLDB 2017 Early Research Contribution Award, IEEE TCDE Early Career Award 2014, CIKM 2017 Best Paper Award, DASFAA 2014 Best Paper Runner-up, and APWeb 2014 Best Paper Award.

**Jiannan Wang** is currently working as an assistant professor at the School of Computer Science in Simon Fraser University, Canada. His research is focused on data cleaning, interactive analytics, and crowdsourcing (especially cost control and latency control). Prior to that, he was a postdoc in the AMPLab at UC Berkeley. He obtained his PhD from the Computer Science Department at Tsinghua University. His recent awards include a best demo award at SIGMOD 2016, a Distinguished Dissertation Award from the China Computer Federation (2013), and a Google Ph.D. Fellowship (2011).

**Yudian Zheng** is currently at Twitter, focusing on building large scale online distributed machine learning systems. His main research interests include data analysis, crowdsourced data management and machine learning. He has published more than 20 papers in leading international conferences/journals, including SIGMOD, KDD, VLDB, TKDE, ICDE, etc.

## 6 RELATED PAST TUTORIALS

We have gave a tutorial at SIGMOD 2017 [33] which focuses on crowdsourced data management. Compared with that tutorial, we focus on the fundamental techniques for crowd data mining.

## 7 EQUIPMENT

The tutorial does not require any special equipment. We will use our own laptop for presentation.

## 8 SLIDES DUE AND PREVIOUS WEBSITES

We understand the due of slides is on July 29, 2018. Actually if the proposal is accepted, we will get the slides and website ready before July 2018. Please kindly check our previous tutorials: http://dbgroup.cs.tsinghua.edu.cn/ligl/papers/sigmod17-tutorial-crowd.pdf. We have also written a survey [32] and a book [31] in crowdsourcing.

## 9 VIDEO SNIPPET

Here are two video links of our tutorial talks at SIGMOD 2017 from YouTube.

https://www.youtube.com/watch?v=ADAp7XMGtjw
https://www.youtube.com/watch?v=-45JkIVYhvo